\documentclass[preprint]{pasj00} 
\usepackage{lscape}
\usepackage[normalem]{ulem}

\newcommand{\asca}{{\it ASCA}}

\newcommand{\suzaku}{{\it Suzaku}}
\newcommand{\bepposax}{{\it BeppoSAX}}

\newcommand{\xmm}{{\it XMM-Newton}}

\newcommand{\logn}{log $N$ - log $S$ relation}
\newcommand{\Logn}{log $N$ - log $S$ Relation}

\newcommand{\erg}{erg s$^{-1}$}
\newcommand{\ergs}{erg cm$^{-2}$ s$^{-1}$}

\newcommand{\nh}{$N_{\rm H}$}
\newcommand{\lx}{$L_{\rm X}$}

\newcommand{\cosmo}{($H_0$, $\Omega_{\rm m}$, $\Omega_{\lambda}$)}

\begin{document}
\SetRunningHead{Y. Ueda et al.}
{Revisit of Local AGN X-ray Luminosity Function 
with MAXI}
\Received{}
\Accepted{}
\Published{}

\title{
Revisit of Local X-ray Luminosity Function 
of Active Galactic Nuclei with the MAXI Extragalactic Survey
}

\author{
Yoshihiro \textsc{Ueda},\altaffilmark{1}
Kazuo \textsc{Hiroi},\altaffilmark{1}
Naoki \textsc{Isobe},\altaffilmark{1,2}
Masaaki \textsc{Hayashida},\altaffilmark{1}
Satoshi \textsc{Eguchi},\altaffilmark{1,3}
Mutsumi \textsc{Sugizaki},\altaffilmark{4}
Nobuyuki \textsc{Kawai},\altaffilmark{5}
Hiroshi \textsc{Tsunemi},\altaffilmark{6}
Tatehiro \textsc{Mihara},\altaffilmark{4}
Masaru \textsc{Matsuoka},\altaffilmark{4,7}
Masaki \textsc{Ishikawa},\altaffilmark{8}
Masashi \textsc{Kimura},\altaffilmark{6}
Hiroki \textsc{Kitayama},\altaffilmark{6}
Mitsuhiro \textsc{Kohama},\altaffilmark{7}
Takanori \textsc{Matsumura},\altaffilmark{9}
Mikio \textsc{Morii},\altaffilmark{5}
Yujin E. \textsc{Nakagawa},\altaffilmark{10}
Satoshi \textsc{Nakahira},\altaffilmark{4}
Motoki \textsc{Nakajima},\altaffilmark{11}
Hitoshi \textsc{Negoro},\altaffilmark{12}
Motoko \textsc{Serino},\altaffilmark{4}
Megumi \textsc{Shidatsu},\altaffilmark{1}
Tetsuya \textsc{Sootome},\altaffilmark{4}
Kousuke \textsc{Sugimori},\altaffilmark{5}
Fumitoshi \textsc{Suwa},\altaffilmark{12}
Takahiro \textsc{Toizumi},\altaffilmark{5}
Hiroshi \textsc{Tomida},\altaffilmark{7}
Yohko \textsc{Tsuboi},\altaffilmark{9}
Shiro \textsc{Ueno},\altaffilmark{7}
Ryuichi \textsc{Usui},\altaffilmark{5}
Takayuki \textsc{Yamamoto},\altaffilmark{4}
Kazutaka \textsc{Yamaoka},\altaffilmark{13}
Kyohei \textsc{Yamazaki},\altaffilmark{9}
Atsumasa \textsc{Yoshida},\altaffilmark{13}
and the MAXI team} 
\altaffiltext{1}{Department of Astronomy, Kyoto University, Oiwake-cho, Sakyo-ku, Kyoto 606-8502}
\email{ueda@kusastro.kyoto-u.ac.jp}
\altaffiltext{2}{Institute of Space and Astronautical Science (ISAS),
Japan Aerospace Exploration Agency (JAXA), 3-1-1 Yoshino-dai, Chuo-ku, Sagamihara, Kanagawa 252-5210}
\altaffiltext{3}{National Astronomical Observatory of Japan, 2-21-1, Osawa, Mitaka City, Tokyo 181-8588}
\altaffiltext{4}{MAXI team, Institute of Physical and Chemical Research (RIKEN), 2-1 Hirosawa, Wako, Saitama 351-0198}
\altaffiltext{5}{Department of Physics, Tokyo Institute of Technology, 2-12-1 Ookayama, Meguro-ku, Tokyo 152-8551}
\altaffiltext{6}{Department of Earth and Space Science, Osaka University, 1-1 Machikaneyama, Toyonaka, Osaka 560-0043}
\altaffiltext{7}{ISS Science Project Office, Institute of Space and Astronautical Science (ISAS), Japan Aerospace Exploration Agency (JAXA), 2-1-1 Sengen, Tsukuba, Ibaraki 305-8505}
\altaffiltext{8}{School of Physical Science, Space and Astronautical Science, The graduate University for Advanced Studies (Sokendai), Yoshinodai 3-1-1, Chuo-ku, Sagamihara, Kanagawa 252-5210}
\altaffiltext{9}{Department of Physics, Chuo University, 1-13-27 Kasuga, Bunkyo-ku, Tokyo 112-8551}
\altaffiltext{10}{Research Institute for Science and Engineering, Waseda University, 17 Kikui-cho, Shinjuku-ku, Tokyo 162-0044}
\altaffiltext{11}{School of Dentistry at Matsudo, Nihon University, 2-870-1 Sakaecho-nishi, Matsudo, Chiba 101-8308}
\altaffiltext{12}{Department of Physics, Nihon University, 1-8-14 Kanda-Surugadai, Chiyoda-ku, Tokyo 101-8308}
\altaffiltext{13}{Department of Physics and Mathematics, Aoyama Gakuin University,\\ 5-10-1 Fuchinobe, Chuo-ku, Sagamihara, Kanagawa 252-5258}


\maketitle

\begin{abstract}

We construct a new X-ray (2--10 keV) luminosity function of
Compton-thin active galactic nuclei (AGNs) in the local universe,
using the first MAXI/GSC source catalog surveyed in the 4--10 keV
band. The sample consists of 37 non-blazar AGNs at
$z=0.002-0.2$, whose identification is highly ($>97\%$) complete. We
confirm the trend that the fraction of absorbed AGNs with \nh\ $>
10^{22}$ cm$^{-2}$ rapidly decreases against luminosity (\lx), from
0.73$\pm$0.25 at \lx\ = $10^{42-43.5}$ erg s$^{-1}$ to 0.12$\pm0.09$
at \lx\ = $10^{43.5-45.5}$ erg s$^{-1}$. The obtained luminosity
function is well fitted with a smoothly connected double power-law
model whose indices are $\gamma_1 = 0.84$ (fixed) and $\gamma_2 =
2.0\pm0.2$ below and above the break luminosity, $L_{*} =
10^{43.3\pm0.4}$ ergs s$^{-1}$, respectively.  While the result of
the MAXI/GSC agrees well with that of HEAO-1 at \lx\ $\gtsim 10^{43.5}$ erg
s$^{-1}$, it gives a larger number density at the lower luminosity
range. Comparison between our luminosity function in the 2--10 keV
band and that in the 14--195 keV band obtained from the Swift/BAT
survey indicates that the averaged broad band spectra in the 2--200
keV band should depend on luminosity, approximated by $\Gamma\sim1.7$
for \lx\ $\ltsim 10^{44}$ erg s$^{-1}$ while $\Gamma\sim 2.0$ for \lx\
$\gtsim 10^{44}$ erg s$^{-1}$.  This trend is confirmed by the
correlation between the luminosities in the 2--10 keV and 14--195 keV
bands in our sample. We argue that there is no contradiction in the
luminosity functions between above and below 10 keV once this effect is
taken into account.

\end{abstract}

\section{INTRODUCTION}
\label{sec:INTRODUCTION}

The tight correlation between the mass of a supermassive black hole
(SMBH) in a galactic center and that of the budge found in the local
universe (\cite{1998AJ....115.2285M}; \cite{2000ApJ...539L...9F};
\cite{2000ApJ...539L..13G}; \cite{2003ApJ...589L..21M};
\cite{har04}; \cite{hop07}; \cite{kor09}; \cite{gul09}) leads to
an idea of the ``co-evolution'' of SMBHs and galaxies.  Thus,
understanding the growth of SMBHs is a fundamental issue to elucidate
the cosmological history of the universe in the relation with the
evolution of galaxies. The key objects to study this are active
galactic nuclei (AGNs), the phenomena where the SMBH gains its mass by
accreting gas.

The most basic observational quantities to describe the cosmological
evolution of AGNs is the luminosity function (LF), the number density
per comoving space as a function of luminosity and redshift. To derive
the AGN LF, a statistically well-defined sample (unusually, a flux
limited one) with complete identification obtained by unbiased surveys
is required. Hard X-ray observations at energies above a few keV provide
the most efficient and complete surveys to detect the whole AGNs
including obscured ones (so-called ``type 2'' AGNs), the major class in
this population (e.g., see \cite{gil07}), thanks to its strong
penetrating power against absorption by the surrounding material and to
little contamination from stars in the host galaxy. In the past several
years, combinations of hard X-ray surveys above 2 keV with different
survey depths have revealed the evolution of LF of AGNs constituting the
major part of the Cosmic X-ray background (CXB), which gives strong
constraints on the scenario of the SMBH growth from $z=0$ to $z\sim5$
(e.g., \cite{ued03}; \cite{laf05}; \cite{bar05}; \cite{sil08};
\cite{ebr09}; \cite{yen09}; \cite{air10}).

To establish the X-ray LF of AGNs in the local universe is of great
importance among these efforts, since it gives the reference for any
evolution models. While \citet{ued03}, \citet{sil08}, \citet{ebr09}, and
\citet{yen09} have found that the luminosity dependent density evolution
(LDDE; this term was originally introduced by \cite{miy00}) best
describes of the X-ray AGN LF above 2 keV, \citet{air10} recently
suggest that the luminosity and density evolution (LADE) where the
``shape'' of the LF is constant over the whole redshift range gives a
similarly good fit to their data. \citet{ued03} employ the local
AGN LF in the 2--10 keV band based on the HEAO-1 all sky survey, by
using the sample consisting of 49 AGNs compiled by \citet{shi06}.
\citet{saz04}
\footnote{The AGN X-ray fluxes (and consequently luminosities) 
used by \citet{saz04} were underestimated by a
factor of 1.4 due to an error in the count rate to flux conversion
\citep{saz08}. In this paper, we correct for this error whenever we
refer to the results of \citet{saz04}.}  has also determined the local
AGN LF in the 3--20 keV band from the RXTE/Slew survey \citep{rev04},
whose integrated volume emissivity corrected for incompleteness is found
to be by a factor of $\sim$2 smaller than the HEAO-1 result converted
into the same energy band, however.

More recently, hard X-ray surveys above 10 keV performed by the Swift
and INTEGRAL satellites also determine the local AGN LF in the 14--195
keV or 15--55 keV band (Swift; \cite{tue08}; \cite{bur11}) and in
the 20--40 keV or 17--60 keV band (INTEGRAL; \cite{bec06b};
\cite{saz07}), respectively. The advantage of these surveys is the
least biases against heavily obscured AGNs, although the observed
fraction of Compton thick objects with an absorption column density of
\nh\ $>10^{24}$ cm$^{-2}$ is found to be as small as $<$5
percent in the total sample (\cite{tue08}; \cite{bur11}). It is
found the shape of the AGN LF above 10 keV as determined by Swift/BAT
looks significantly different from the \citet{shi06} result if the
luminosity is simply converted into the other band by assuming a
typical AGN spectrum (characterized by a power law with a photon index
of $\approx$1.8.). The reasons of this discrepancy have not been
understood yet.

Thus, it is timely to revisit the local X-ray AGN LF below 10 keV from
a new survey independently, in order to check the consistency with the
previous works and solve the apparent contradictions among them. The
Monitor of All-sky X-ray Image (MAXI) mission on the International
Space Station \citep{mat09}, currently in orbit, provides a valuable
opportunity for this. \citet{hir11} produce the first source catalog
of the MAXI/Gas Slit Camera (GSC; \cite{mih11}; \cite{sug11}) at high
galactic latitudes ($|b|>10^\circ$), by compiling the data in the
4--10 keV band accumulated for the first 7 month since the start of
its nominal operation. The catalog contains 51 AGNs detected
with a significance above 7$\sigma$ consisting of 39 Seyfert
galaxies and 12 blazars. In this paper, we constrain the local AGN LF
in the 4--10 keV band by using only non-blazar AGNs (i.e.,
Seyferts) in the \citet{hir11} catalog. We also determine the
intrinsic distribution of absorption column density of AGNs (so-called
\nh\ function) in the local universe using the same sample, and
compare it with the previous results. Section~2 briefly describes the
source sample and their X-ray spectral properties in terms of an
absorption column density and a photon index determined from various
observatories. Section~3 describes the analysis method and obtained
results.  We discuss the implication in Section~4. The cosmological
parameters of \cosmo\ = (70$h_{70}$ km s$^{-1}$ Mpc$^{-1}$, 0.3, 0.7)
are adopted throughout the paper. The ``log'' symbol represents the
base-10 logarithm, while ``ln'' the natural logarithm.

\section{Sample}

To investigate the local LF and \nh\ function of AGNs, we collect the
37 non-blazar AGNs from the \citet{hir11} catalog at
$z=0.002-0.2$ that constitute a statistically unbiased sample detected
in the 4--10 keV band from an area of 34,000 deg$^2$. Here we
exclude Cen A, located at $z<0.002$, and ESO~509--066, which has
double nuclei and may be contaminated by nearby sources (see Table~1
in \cite{hir11}).  The four ``confused'' sources are ignored even if
they contain contribution from AGNs (like NGC 6814), which do not
affect our results. As noted in \citet{hir11}, the list of the X-ray
brightest AGNs in the all sky has significantly changed from that of
the HEAO-1 survey performed 30 years ago; among 39 MAXI/GSC
detected AGNs, only 17 objects are listed in both sample by
\citet{pic82} and that used by \citet{shi06}.  The flux limit of the
MAXI sample corresponds to $1.5\times10^{-11}$ \ergs\ (1.2 mCrab) in
the 4--10 keV band. Figure~1 shows the \logn\ (integral form) of these
AGNs in the 4--10 keV band, obtained by using the area curve presented
in Figure~9 of \citet{hir11}.

\begin{figure}[htbp]
  \begin{center}
    \FigureFile(80mm,80mm){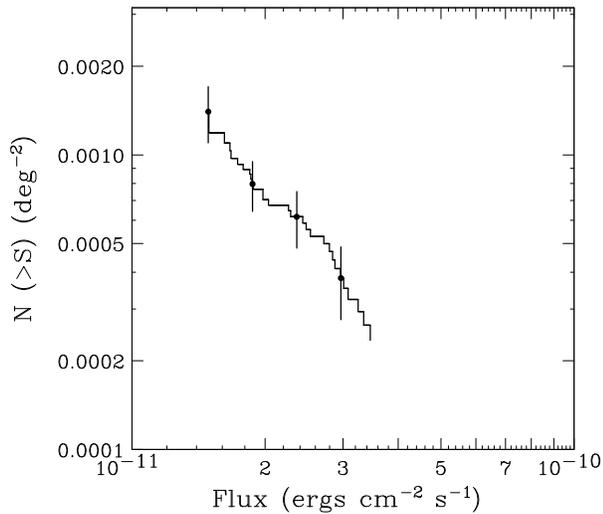}
  \end{center}
  \caption{\Logn\ of non-blazar AGNs in the 4--10 keV band
used in our analysis, determined from the first MAXI/GSC source catalog 
by \citet{hir11}. }
  \label{fig1}
\end{figure}

Table~1 summarizes the AGN list, where the first to sixth columns
represent the catalog source No., MAXI source name, counterparts,
optical type, 4--10 keV flux, and redshift, respectively. Although it is
known that using the spectroscopic redshifts to estimate the distance of
very nearby objects is subject to uncertainties due to the galaxy proper
motion, we adopt these values for consistency with the analysis of the
Swift/BAT AGNs in \citet{tue08}. We confirm that even if we instead
adopt redshifts corrected for the infall into the Virgo cluster by
\citet{mou00} to calculate the luminosity of AGNs at $z<0.01$, our
results of both LF and \nh\ function are little affected. Here we only
distinguish between two optical classes, ``AGN1'' (Seyfert 1.0--1.5) or
``AGN2'' (Seyfert 1.8--2.0), for simplicity. We can regard that this AGN
sample is nearly complete (99.3\%), because 142 out of the total 143
X-ray sources are identified there. The flux errors due to the
statistical uncertainties are better than $14\%$ for all the objects,
and hence they are not taken into account in the following analysis.

To compare our result with the previous works easily, we construct the
AGN LF in the ``intrinsic'' 2--10 keV luminosity corrected for the
absorption (i.e., before absorption) at the source frame (hereafter
represented as \lx). Since we have the count rate in the 4--10 keV
band from the MAXI/GSC survey, it is necessary to convert it to \lx\
by using the spectral information as well as the redshift for each
source. Fortunately, we are able to find results of spectral fits in
the 0.2--10 (or 0.5--10) keV band in the literature for 33 (out of 37)
AGNs, which were obtained from data of either \asca, \xmm, \bepposax,
Swift/XRT, or \suzaku. The spectral quality is sufficiently good in
most cases, and hence we neglect their errors in the following
analysis. The best-fit photon index $\Gamma$, absorption column
density \nh\ (at the source frame), and calculated luminosity from
these parameters (\lx) are listed in the 7th to 9th columns of
Table~1, respectively, together with the reference for the spectral
parameters (10th column). In the conversion from the MAXI/GSC count
rate into \lx, we consider a reflection component from cold, optically
thick matter \citep{mag95} with a solid angle of $\Omega=2\pi$ as
adopted in \citet{ued03}, although this does not affect our result of
the LF. For the remaining four targets\footnote{IRAS~05078+1626, 2MASX
J09235371--3141305, 4C~+18.51, 1RXS J213623.1--622400}, we perform the
same image analysis of the MAXI/GSC data in the 2--4 keV band as that
in \citet{hir11} to obtain the hardness ratio between the 2--4 keV and
4--10 keV count rates. We first calculate the corresponding photon
index $\Gamma$ without considering any absorption; if we obtain
$\Gamma<1.9$, then we derive an absorption column density at the
source redshift assuming an intrinsic power law with $\Gamma=1.9$. The
results of $\Gamma$ and \nh\ with $1\sigma$ statistical errors
estimated in this way are also listed in Table~1 for these 4 targets.
Figure~2 shows the redshift ($z$) versus luminosity (\lx) plot for our
sample. The open and filled circles correspond to those with a column
density of log \nh\ $< 22$ and log \nh\ $< 24$, respectively. The
optical type-2 AGNs are further marked with the diagonal crosses.

\begin{figure}[htbp]
  \begin{center}
    \FigureFile(80mm,80mm){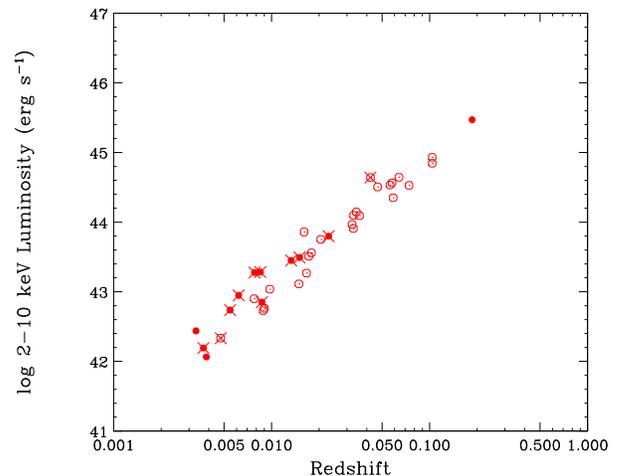}
  \end{center}
  \caption{Redshift versus luminosity plot of our sample.
The luminosity is the intrinsic (before absorption) one in the 2--10 keV
 band estimated from the 4--10 keV MAXI/GSC count rate corrected for 
an absorption column density. The open and filled circles correspond 
to the X-ray unabsorbed (log \nh\ $<22$) and absorbed AGNs, respectively.
Optical type 2 AGNs are marked with the diagonal crosses.
}
  \label{fig2}
\end{figure}

\section{Analysis and Results}

\subsection{Analysis Method}

Our goal is to determine both the \nh\ function and absorption-corrected
2-10 keV LF of X-ray AGNs in the local universe. The calculation follows
the same procedure as presented in \citet{ued03}, to which we
refer the reader for details. The same notation convention is adopted in
this paper. The \nh\ function, $f(L_{\rm x}, z; N_{\rm H})$ $d$log \nh,
represents a probability of finding an AGN with an absorption column
density between log \nh\ and log \nh\ + $d$log \nh\ at a given
luminosity, \lx, and redshift, $z$. For convenience, we assign log \nh\
= 20 for AGNs without any significant absorption, and consider only the
range of log \nh\ $\leq$ 24, since no Compton thick AGNs are present in
the current sample. It is normalized as
\begin{equation}
\int_{20}^{24} f (L_{\rm X}, z; N_{\rm H}) d{\rm log} N_{\rm H} = 1.
\end{equation}
The LF, $\Phi (L_{\rm X}, z)$ in units of Mpc$^{-3}$, is defined so
that
$
        d \Phi (L_{\rm X}, z)/d{\rm log} L_{\rm X}
$
gives the co-moving space density of all (Compton-thin) AGNs in a
luminosity range between log \lx\ and log \lx\ + $d$log \lx\ at a
redshift of $z$.

From the list of \nh\ and \lx\ in our sample, the best-fit parameters
are searched for by minimizing the likelihood estimator defined as
\begin{equation}
\label{eq1}
L = -2 \sum_{i} {\rm ln}
\frac{N(N_{{\rm H}i}, L_{{\rm X}i}, z_i)}
{\int \int \int N( N_{\rm H}, L_{{\rm X}}, z) d{\rm log} N_{\rm H} d{\rm log} L_{\rm X} d z},
\label{eq-likelihood}
\end{equation}
where the suffix $i$ denotes each object. The term 
$N(N_{\rm H}, L_{{\rm X}}, z)$ represents the expected number 
from the survey, 
\begin{eqnarray}
\nonumber &&N(N_{\rm H}, L_{{\rm X}}, z) = 
f(L_{\rm X}, z; N_{\rm H}) \times \\
&&\;\;\;\;\;
\frac{d \Phi (L_{\rm X}, z)}{d{\rm log} L_{\rm X}}
d_{\rm A}(z)^2 (1+z)^3 c \frac{d\tau}{dz}(z) A (N_{\rm H},L_{\rm X},z), 
\label{eq-likelihood2}
\end{eqnarray}
where $d_{\rm A}(z)$ is the angular distance, $d\tau/d z$ 
the differential look back time, and $A (N_{\rm
H},L_{\rm X},z)$ the survey area, given as a function of flux 
that is calculated from $N_{\rm H}$, $L_{\rm X}$, and $z$.
The minimization process is carried out on the MINUIT software
package. The $1\sigma$ error for a single parameter can be estimated
from the parameter range that increases the $L$ value by 1.
The fit applied to the unbinned data here cannot estimate the
normalization of the LF. Hence, we determine it so that the expected
source number agrees with the observed one, and estimate its relative
uncertainty only from its Poisson error ($1/\sqrt{N}$, where $N=37$).

\subsection{\nh\ Function}

To avoid coupling between the \nh\ function and LF, we determine them
step by step in the same way as \citet{ued03}, considering the small
sample size.  First, we constrain the \nh\ function by adopting the
``delta-function'' approximation for the LF only based on the sample
list. It reduces the formula~(2) to a simpler form (see equation (6) in
\cite{ued03}) where the intrinsic \nh\ distribution can be evaluated
directly from its observed \nh\ histogram by taking into account the
\nh\ dependence of the survey area. For the four sources whose
absorptions are estimated from the hardness ratios of the MAXI data and
hence have non-negligible statistical errors, we take into account the
uncertainties in \nh\ by introducing the ``\nh\ response matrix
function'' as done in \citet{ued03} (see their Section~4.1).

As for the shape of the \nh\ function, we adopt a modified version of
that used in \citet{ued03}. The difference from \citet{ued03} is that
(1) we allow such a case that the \nh\ function at log \nh\ $<22$ is
smaller than that at log \nh\ $>22$, (2) we assign 4 discrete bins
with the same width between log \nh = 20--24 for simplicity,
considering the practical difficulties to determine \nh\ with an
accuracy better than $d$log \nh\ $< 1$ for objects without good X-ray
spectral data. We define the absorption fraction $\psi (L_{{\rm X}})$
as that of AGNs with log \nh\ = 22--24 among those with log \nh\ =
20--24, which is given as a function of luminosity.  Its possible
redshift dependence is ignored, because our sample consists of only
local AGNs. The form of the \nh\ function is expressed differently for
two $\psi (L_{{\rm X}})$ ranges;
\begin{eqnarray}
\label{eq-former}
\nonumber &&({\rm for}\;\; \psi (L_{\rm X}) < \frac{1+\epsilon}{3+\epsilon})\;\;\;\;\; \\
&&f(L_{{\rm X}}, z; N_{{\rm H}}) = 
\small
\begin{tabular}{ll}
$1-\frac{2+\epsilon}{1+\epsilon}\psi (L_{{\rm X}})$ & $(20 \leq {\rm log} N_{\rm H} < 21)$\\
$\frac{1}{1+\epsilon} \psi (L_{{\rm X}})$ & $(21 \leq {\rm log} N_{\rm H} < 22)$\\
$\frac{1}{1+\epsilon} \psi (L_{{\rm X}})$ & $(22 \leq {\rm log} N_{\rm H} < 23)$\\
$\frac{\epsilon}{1+\epsilon} \psi (L_{{\rm X}})$ & $(23 \leq {\rm log} N_{\rm H} < 24).$\\
\end{tabular}
\normalsize
\end{eqnarray}
and
\begin{eqnarray}
\label{eq-latter}
\nonumber &&({\rm for}\;\; \psi (L_{\rm X}) \geq \frac{1+\epsilon}{3+\epsilon})\;\;\;\;\; \\
&&f(L_{{\rm X}}, z; N_{{\rm H}}) =
\small
\begin{tabular}{ll}
$\frac{2}{3}-\frac{3+2\epsilon}{3+3\epsilon}\psi (L_{{\rm X}})$ & $(20 \leq {\rm log} N_{\rm H} < 21)$\\
$\frac{1}{3}-\frac{\epsilon}{3+3\epsilon}\psi (L_{{\rm X}})$ & $(21 \leq {\rm log} N_{\rm H} < 22)$\\
$\frac{1}{1+\epsilon} \psi (L_{{\rm X}})$ & $(22 \leq {\rm log} N_{\rm H} < 23)$\\
$\frac{\epsilon}{1+\epsilon} \psi (L_{{\rm X}})$ & $(23 \leq {\rm log} N_{\rm H} < 24).$\\
\end{tabular}
\normalsize
\end{eqnarray}
Here $\epsilon$ defines the ratio of the \nh\ function in log
\nh\ = 23--24 to that in log \nh\ = 22--23. It is fixed at 1.3
(instead of 1.7 as adopted in \cite{ued03}), according to the
observed \nh\ distribution in the Swift/BAT 9-month survey (23/18,
\cite{tue08}), 
which agrees with the more recent result by \citet{bur11}
. 
In the former case (equation~\ref{eq-former}), the \nh\ function is flat
above log \nh\ = 21, while in the latter case
(equation~\ref{eq-latter}), the value in log \nh\ = 21--22 is taken to
be the mean of those at log \nh\ = 20--21 and log \nh\ = 22--23. The
maximum absorption fraction is $\psi_{max} =
\frac{1+\epsilon}{3+\epsilon}$, corresponding to the case of $f(N_{{\rm
H}})$ = 0 at log \nh\ = 20--21.

Figure~2 clearly shows that X-ray absorbed AGNs are mostly found in
the lower luminosity range (log \lx\ $< 44$). This confirms the
trend found in many previous works (e.g., \cite{ued03}; \cite{has08})
that the absorption fraction reduces with an increase of the AGN
luminosity. Thus, following \citet{ued03}, we model the absorption
fraction by a linear function of log \lx\ within the maximum (see
above) and minimum values, which is taken to be 0.1.  It is
represented as
\begin{equation}
\psi(L_{{\rm X}}) = {\rm min} [\psi_{\rm max}, {\rm max} [\psi_{44} - \beta ({\rm log} L_{\rm X} - 44), 0.1] ], 
\end{equation}
where $\beta$ and $\psi_{44}$ are the free parameters to be determined
through the likelihood fit.

Table~2 summarizes the best-fit parameters of the \nh\ function and
their $1\sigma$ errors. Figure~3(a) plots the ``intrinsic'' \nh\
function (corrected for the observation bias) for the total sample
(upper), that for low luminosities of log \lx\ $< 43.5$ (middle), and
that for log \lx\ $>43.5$ (lower). The dependence of the absorption
fraction on the luminosity is obvious. The best-fit model of the \nh\
function calculated at the mean \lx\ value in each region is
overplotted. Figure~3(b) shows the ``observed'' histogram of \nh\ for
these 3 luminosity ranges, on which those predicted from the best-fit 
model are superposed.

\begin{figure}[htbp]
  \begin{center}
    \FigureFile(80mm,80mm){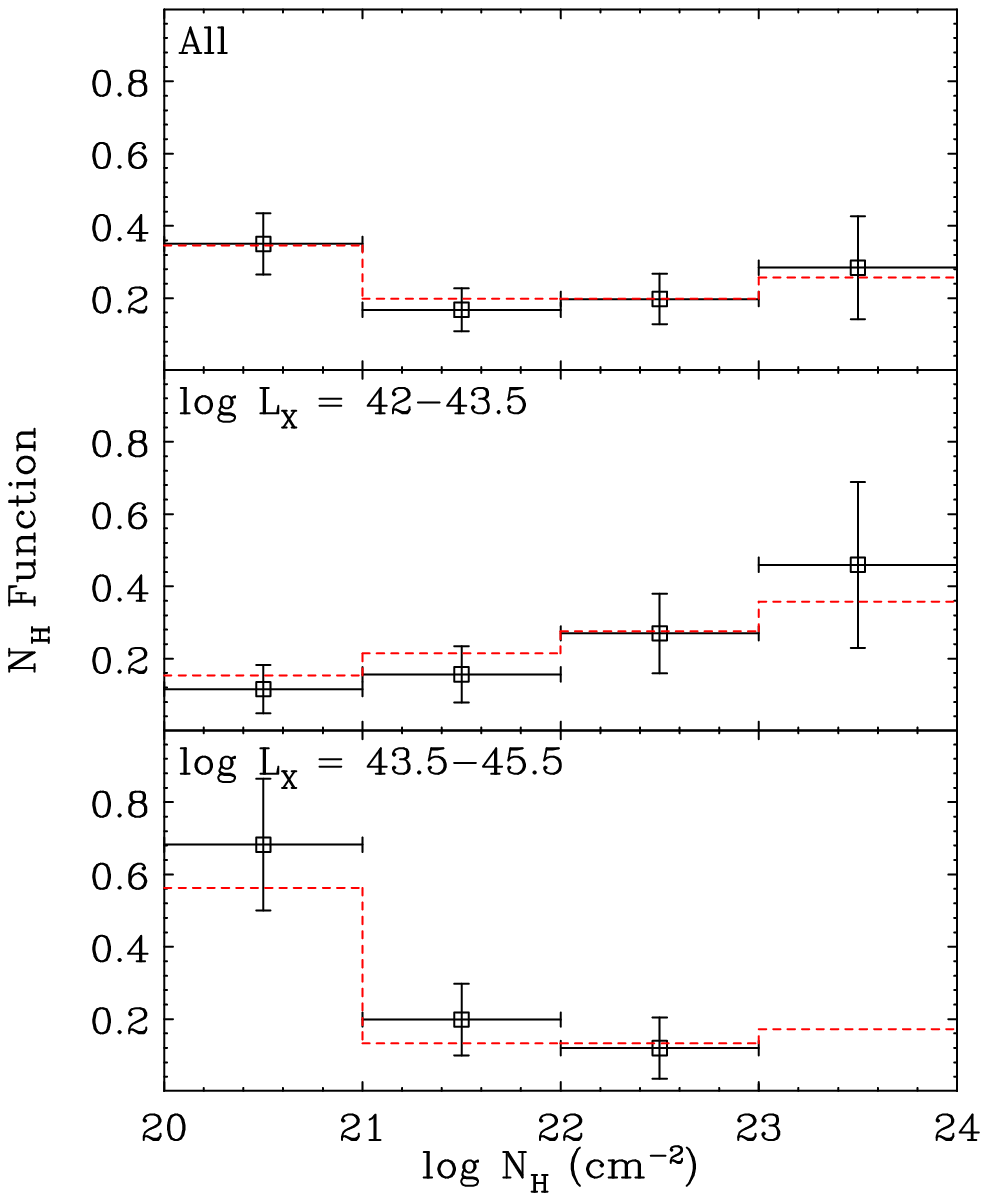}
    \FigureFile(80mm,80mm){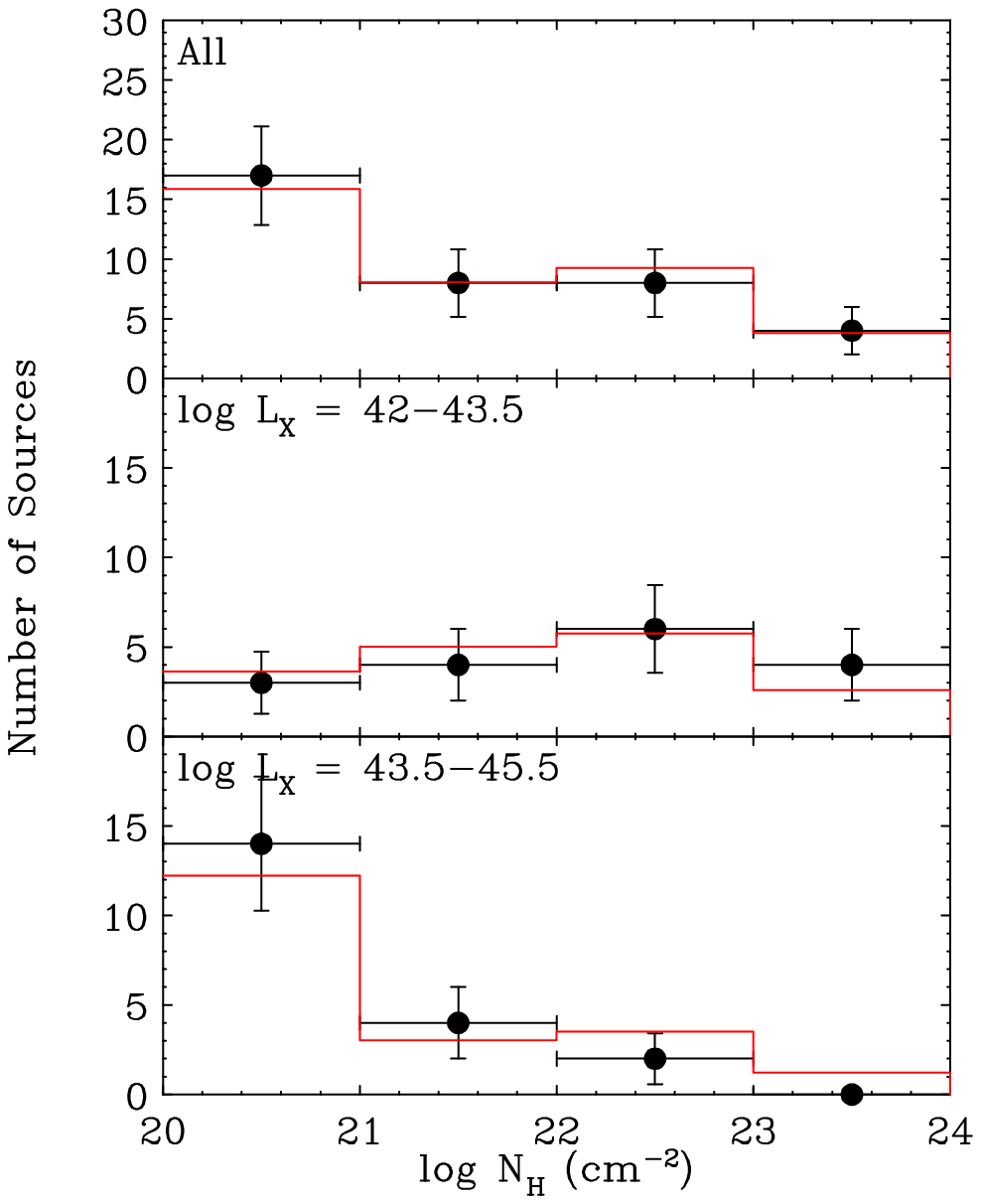}
  \end{center}
  \caption{
(Upper) (a) Intrinsic distribution of absorption column density (\nh
 function). From the upper to lower panels, for the total sample, 
for AGNs with log \lx\ = 42--43.5, and for AGNs with log \lx\ = 43.5--45.5.
The $1\sigma$ error is attached to the data points (open
 circle). The lines (red) are the best-fit \nh\ function calculated at the 
mean luminosity in each luminosity range.
(Lower) (b) The observed histograms of \nh\ without any correction for the
 observational bias. The $1\sigma$ error (Poisson error) is attached to
 the data point (filled circle). The lines (red) show the predicted
 numbers in each log \nh\ bin calculated from the best-fit \nh\
 function.
}
  \label{fig3}
\end{figure}

\subsection{Luminosity Function}

Using the \nh\ function obtained above, we finally determine the
local AGN LF by maximum-likelihood fit according to the
formula~(2). We adopt
the smoothly connected double power law model, one of
the most standard descriptions for X-ray AGN LFs, given as
\begin{equation}
\frac{d \Phi (L_{\rm X}, z=0)}{d{\rm log} L_{\rm X}} 
= A [(L_{\rm X}/L_{*})^{\gamma_1} + (L_{\rm X}/L_{*})^{\gamma_2}]^{-1}.
\end{equation}
To implement the effect of the cosmological evolution, we introduce the 
evolution factor represented by $(1+z)^{p1}$, 
\begin{equation}
\frac{d \Phi (L_{\rm X}, z)}{d{\rm log} L_{\rm X}} 
= \frac{d \Phi (L_{\rm X}, 0)}{d{\rm log} L_{\rm X}} (1+z)^{p1}, 
\label{eq-PDE}
\end{equation}
where we fix $p1=4.2$ based on the result obtained for the LDDE model
in \citet{ued03}.  Note that at $z<0.2$ and log \lx\ $> 42$, their
LDDE model is identical with the pure density evolution model as
represented above.

Due to the limited sample size, we find it difficult to 
constrain the three free parameters of the LF, $\gamma_1$, $\gamma_2$,
and $\L_{*}$ simultaneously. Hence, we fix the power law slope in the low luminosity
range at three different values, $\gamma_1 =1.0$, $\gamma_1 =0.84$
(the best-fit obtained from the Swift/BAT 9-month survey in
\cite{tue08}), and $\gamma_1 = 0.62$ (the best-fit from the LADE model
in \cite{air10}). The results of the likelihood fit for these three
cases are summarized in Table~2. Figure~4 plots the best fit local LF
determined from the MAXI survey for the case of $\gamma_1=0.84$ (black
curve). The data points are calculated by the $N_{\rm data}/N_{\rm
model}$ method \citep{miy01}, to which the $1\sigma$ statistical
errors are attached according to the formula by \citet{geh86}. We find
the local AGN emissivity in the 2--10 keV band integrated over the log
\lx\ = 41--47 range is $W_{\rm 2-10} = (1.37\pm0.23) \times 10^{39}$
$h_{70}$ erg s$^{-1}$ Mpc$^{-3}$ for $\gamma_1 = 0.84$. This value is
close to that obtained by \citet{shi06}, $W_{\rm 2-10} = (9.3\pm1.3)
\times 10^{38}$ $h_{70}$ erg s$^{-1}$ Mpc$^{-3}$,
\footnote{
It is larger than that presented in Section 6.1 of
\citet{shi06}, $(5.85\pm1.17)\times 10^{38}$ $h_{70}$ erg s$^{-1}$
Mpc$^{-3}$. This is because while
\citet{shi06} calculated the volume emissivity using observed (i.e.,
absorbed) fluxes of each AGN at luminosity range of log \lx\ $>42$,
we here calculate it for intrinsic (de-absorbed) luminosities by
integrating the analytical expression of the LF down to log \lx\ $=41$.
}
but significantly larger than that by \citet{saz04}, $W_{\rm 2-10} =
(5.2\pm0.5) \times 10^{38}$ $h_{70}$ erg s$^{-1}$ Mpc$^{-3}$,
converted from the 3--20 keV band LF by assuming a power law photon
index of 1.7.

\begin{figure}[htbp]
  \begin{center}
    \FigureFile(80mm,80mm){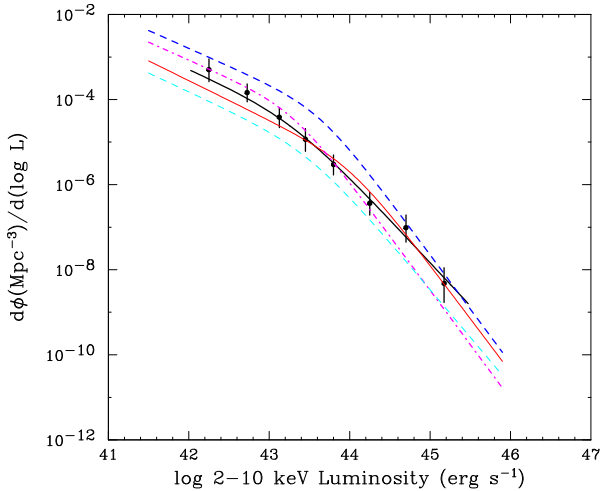}
  \end{center}
  \caption{Local X-ray AGN luminosity function in the 2--10 keV band.
The cosmological parameters of \cosmo\ = (70$h_{70}$ km s$^{-1}$
Mpc$^{-1}$, 0.3, 0.7) are adopted. Data points (black) are the
observed values from the MAXI survey based on the ``$N_{\rm
data}/N_{\rm model}$'' estimator with error bars at $1\sigma$ 
level. The thick solid curve (black) is the best-fit model with
$\gamma_1 = 0.84$ (fixed).  The thin solid curve (red) and thin dashed
curve (cyan) are the results from the HEAO-1 survey by \citet{shi06}
and from the RXTE slew survey by \citet{saz04}, respectively. The
latter is converted from the 3--20 keV band to the 2--10 keV band 
by assuming a power law photon index of 1.7 for
our adopted Hubble constant. The thick dashed (blue) and dot-dashed
(magenta) curves are the luminosity functions derived from the
Swift/BAT 9 month survey \citep{tue08} converted from the 14--195 keV
band to 2--10 keV band by assuming a photon index of 2.0 and
1.7, respectively. }
\label{fig4}
\end{figure}

\section{Discussion}

We revisit the local X-ray luminosity function of non-blazar AGNs,
together with the absorption distribution function, based on the first
source catalog of the on-going MAXI/GSC extragalactic survey. 
In spite of the fact that the current MAXI/GSC source sample is
smaller than those from the HEAO-1 (49 AGNs in \cite{shi06}) and RXTE
(76 AGNs in \cite{saz04}) all sky surveys performed in similar energy
bands, it has some advantages to firmly establish the statistical
properties of AGNs below 10 keV in the following points.
(1) The sample is highly complete (99.3\% = 142/143, or $97$\% = 37/38
in the worst case). (2) Since we have adopted a relatively conservative
threshold in the source selection in our catalog ($7\sigma$), our sample
is less subject to the flux uncertainties in the faintest end ($<14\%$)
and is considered to be free from Eddington's bias as verified in \logn\
(Figure~1). This could actually be a problem in the sample in
\citet{shi06}, who had to correct for such biases by simulation (see
their Appendix). (3) The AGN fluxes are determined from the data
collected from many scans (15 times per day), and hence can be regarded
as the long-term averaged flux, less affected by short term variability
than those obtained from a few snap-shot observations. (4) The energy
band of 4--10 keV band is more suitable for detecting obscured AGNs
(except for Compton thick ones), thus reducing the observation biases
for the \nh\ function determination. It is expected that the MAXI/GSC
4--10 keV sample has intermediate characteristics between the surveys in
softer X-rays than 4 keV, and hard X-ray surveys in above 10 keV.

In fact, we find that the ``observed'' fraction of absorbed AGNs with
log \nh\ $> 22$, 32\% (=12/37), is higher than the HEAO-1 (20\%=10/49)
and RXTE (22\%=17/76) results, while it is lower than that obtained from
the Swift/BAT survey above 15 keV, 49\% (=42/86; \cite{tue08}). We
obtain the intrinsic \nh\ distribution by correcting for the
observational biases that AGNs with heavier absorptions are harder to be
detected due to the reduction of the count rates. The overall shape
combined from both low and high luminosity samples (Figure~3) is well
consistent with the \nh\ distribution obtained from the Swift/BAT
survey, which show an almost flat distribution above log \nh\ $> 21$
with a weak peak in the log \nh\ = 20--21 bin \citep{tue08}. We also
confirm the strong dependence of the absorption fraction on the X-ray
luminosity. We obtain the best-fit formula to describe this relation
slightly different from that in \citet{ued03}, who included much fainter
AGN samples in the analysis. The slope of the absorption fraction with
respect to log \lx\ is steep ($\beta=0.23\pm0.02$ instead of
$\beta=0.10$ in \cite{ued03}), and reaches a higher maximum value ($\psi
= 0.82$ instead of $\psi = 0.57$) at low luminosities. Such sharp (even
sharper) change of the absorption fraction against the luminosity around
log \lx\ $\approx$43.5 is also found in the Swift/BAT sample
(\cite{tue08}; \cite{bur11}). The statistical uncertainty is quite large
at present, however. It is of great importance to investigate redshift
evolutions of the \nh\ function and the relation between the absorption
fraction and luminosity by using much larger samples, which shall be
left for future work.

To compare with the past results obtained from surveys in similar
energy bands, we overlay the best-fit LFs obtained by \citet{shi06} and
\citet{saz04} with the thin solid (red) and thin dashed (cyan) curves,
respectively, in Figure~4. The RXTE LF is converted from the 3--20 keV
band into the 2--10 keV band by assuming a photon index of 1.7 in our
adopted cosmological parameters ($H_0=70$ km s$^{-1}$ Mpc$^{-1}$ instead
of $H_0=75$ km s$^{-1}$ Mpc$^{-1}$ in \cite{saz04}).  The systematic
uncertainties due to the choice of photon index are small in this
case ($18$\% in the luminosity within a range of 1.7--2.0). Here, the
normalization of the LF is corrected for the maximum factor of
incompleteness ($1/0.7$), assuming that the unidentified targets are all
AGNs whose luminosity and redshift distribution is the same as those of
the identified sample. As already reported by \citet{saz04}, the RXTE
result lies significantly lower than the HEAO-1 results by a factor of
$\sim 2$. The origin for this discrepancy is unclear, but we do not
pursue it further in this paper.

As clearly seen in Figure~4, our MAXI LF is closer to the
HEAO-1 LF than the RXTE LF. In particular, it is in good agreement
with the HEAO-1 result at high luminosity range above log \lx\ = 43.5.
However, the MAXI LF gives a larger number density at lower
luminosities by a factor of $\sim$2--3. 
By assuming a similar slope of the LF in the low luminosity range
($\gamma_1 = 0.84-1.0$), the MAXI LF favors a smaller break
luminosity, log $L_{*}$, 42.9--43.9, than the best-fit HEAO-1 value
(log $L_{*}$ = $44.0\pm0.4$), though within the statistical
errors. The discrepancy can be partially explained if the absorption
fraction at these low luminosity range is underestimated than the reality
in the previous work. In fact, according to the best-fit \nh\
function, the absorption fraction at log \lx\ = 42.5 is estimated to
be $\psi = 0.72\pm0.04$ in our work, while it is $\psi = 0.65$ in
\citet{ued03}. Due to the coupling with the \nh\ function 
in constraining the LF parameters through the maximum likelihood fit (see
equation~(2)), the estimated LF for all AGNs with log \nh\ = 20--24
would become smaller if we assume a lower absorption fraction in the
\nh\ function, because it is hard to detect objects with large column
densities of log \nh\ $>23$ in the 2--10 keV band survey, and its
space density can be only constrained by the extrapolation from the
lower column-density range.

Comparison of our AGN LF in the 2--10 keV band with the hard X-ray
($>10$ keV) LF determined with the Swift/BAT and INTEGRAL surveys
provides insights on the broad band properties of local AGNs. Since the
fraction of Compton-thick AGNs in those hard X-ray surveys are
negligibly small, we can directly compare them with our result obtained
for Compton-thin AGNs. In Figure~4, we also plot the best-fit form of
the LF by \citet{tue08} by converting the luminosity from 14--195 keV to
2--10 keV. In this case, the assumption of the spectrum strongly affects
the result. We adopt two photon index, $\Gamma=1.7$ (thick dot-dashed,
magenta) and $\Gamma=2.0$ (thick dashed, blue). Obviously, the shape of
the LF is not the same between these bands if a single spectrum index is
assumed for all AGNs. At low luminosity range of log \lx\ $\ltsim 44$,
the normalizations of the two LFs become consistent with each other for
$\Gamma \sim 1.7$, while at higher luminosities, the conversion with
$\Gamma \sim 2.0$ gives a better agreement.

This result indicates that the averaged shape of broad band X-ray
spectra of these AGNs depends on the luminosity, in the sense that more
luminous AGNs show a steeper slope in the 2--200 keV range on
average. To confirm this picture, we make the correlation plot of
luminosity between the 2--10 keV and 14--195 keV bands using our MAXI
sample (Figure~5). Here the hard X-ray luminosities are taken from the
Swift/BAT 22-month catalog except for two AGNs that are not detected
there; 2MASX J18470283--7831494 for which we refer to the Swift/BAT
58-month catalog, and 4C~+18.51, which is not yet detected by the 58
month data and has only a flux upper limit of $1.1\times10^{-11}$ \ergs\
in the 14--195 keV band. We plot two lines corresponding to a power law
photon index of $\Gamma=1.7$ (solid, magenta) and $\Gamma=2.0$ (dashed,
blue). The trend that the AGNs with lower luminosities have flatter
slope is indeed seen.

\begin{figure}[htbp]
  \begin{center}
    \FigureFile(80mm,80mm){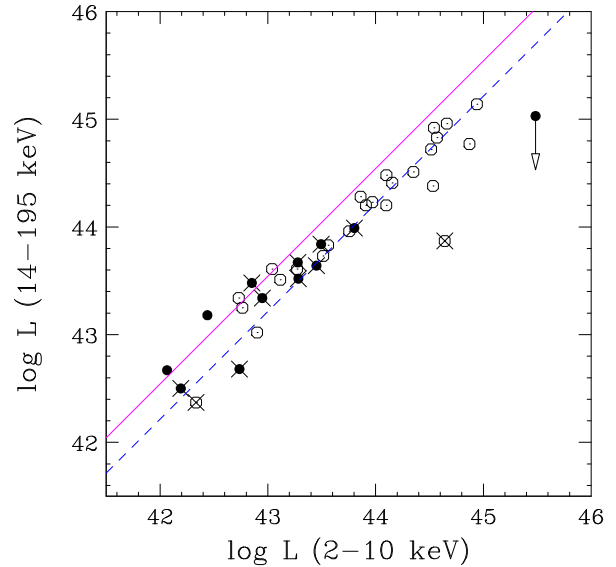}
  \end{center}
  \caption{Luminosity correlation between the 2--10 keV and 14--195
keV bands, determined from the MAXI/GSC and Swift/BAT surveys,
respectively. No K-correction and that for absorption is applied to
the 14--195 keV luminosity.  The open and filled circles are X-ray
unabsorbed (log \nh\ $<22$) and absorbed (log \nh\ $\geq22$) AGNs,
respectively. Optical type 2 AGNs are marked with the diagonal
crosses. The arrow denotes the upper limit for an AGN that is not
detected in the Swift/BAT 58-month survey \citep{bau11}. The dashed
(blue) and solid (magenta) lines correspond to the ratio expected
from a single power law spectrum with a photon index of 2.0 and 1.7,
respectively. }
  \label{fig5}
\end{figure}

We have shown that since the dependence of the averaged X-ray broad band
spectra on luminosity makes the direct comparison of LFs constructed in
different energy bands (below and above 10 keV) not straightforward, its
apparent difference in the LF shape is not ``contradiction''. This
effect must be taken into account when one constructs a LF in a uniform
way by compiling results of X-ray surveys performed in different energy
bands. There are two explanations for the reasons of the luminosity
dependence of the 2--200 keV spectra. Recent studies of nearby AGNs have
suggested that the intrinsic power law components of Seyfert 1 are
steeper than those of Seyfert 2 (e.g., \cite{mal03}; \cite{bec06a};
\cite{tue08}). Because of the strong dependence of absorption fraction
on the luminosity, we mostly detect only type 1 AGNs in the surveys at
the high luminosity range, leading to the trend we see. Second
possibility is the effect of a reflection component, which could be more
significant at lower luminosities, as implied by the ``X-ray Baldwin''
effects \citep{iwa93}. The presence of a reflection hump in the spectra
increases the observed flux in the hard X-ray band, peaked at $\sim 20 $
keV, and hence the apparent slope over the 2--200 keV band becomes
flatter. To distinguish these two effects, systematic studies of the
broad X-ray band spectra of both type 1 and type 2 AGNs at various
luminosity ranges are necessary.

\section{Conclusion}

We have constructed the local AGN X-ray luminosity function, utilizing
our new sample consisting of 37 non-blazar AGNs at $z=0.002-0.2$
detected in the 4--10 keV band from the first MAXI/GSC source catalog
by \citet{hir11}. The sample is highly complete $>97\%$, and is less
subject to uncertainties in the measured fluxes compared with the past
all-sky survey missions above 2 keV. The conclusion of our work is
summarized as follows.

\begin{itemize}

\item We strongly confirm the trend that there exist more absorbed
AGNs at lower luminosities. The fraction of absorbed AGNs with log
\nh\ = 22--24 among those with log \nh\ $< 24$ corrected for the
observational biases changes from 0.73$\pm$0.25 at log \lx\ = 42--43.5
to 0.12$\pm0.09$ at log \lx\ = 43.5--45.5. The estimated absorption
distribution (\nh\ function) is consistent with the Swift/BAT and
INTEGRAL results obtained above 10 keV.

\item The shape of the intrinsic luminosity function of Compton thin
AGNs can be fit with a smoothly connected double power law. For a
fixed slope of $\gamma_1=0.84$ at a lower luminosity range, we obtain
a break luminosity of log $L_{*}$ = 43.3$\pm$0.4 and a higher
luminosity slope of $\gamma_2 = 2.0\pm0.2$. The $L_{*}$ value is
somewhat smaller than the HEAO-1 result. The integrated emissivity
over log \lx\ = 41--47 is found to be $(1.37\pm0.23) \times 10^{39}$
erg s$^{-1}$ Mpc$^{-3}$, which is only slightly larger than the
previous estimate by HEAO-1. The space density agrees with the HEAO-1
result at log \lx\ $gtsim 43.5$ but is larger at the lower luminosity
range. This may be partially explained by the smaller biases against
absorption in our survey in the 4--10 keV band, which lead to a better
estimate of the \nh\ function.

\item We compare our AGN luminosity function in the 2--10 keV band
with those derived above 10 keV, by converting the luminosities by
assuming a single power law spectrum. We find that the space densities
matches with each other for $\Gamma \sim 1.7$ at log \lx\ $<44$, while
they do for $\Gamma \sim 2.0$ at higher luminosities. This suggests
the luminosity dependence of the averaged broad X-ray band spectra
over the $\sim 2-200$ keV band. The trend is indeed confirmed by the
luminosity correlation between the MAXI and Swift/BAT data in our
sample.

\end{itemize}

\bigskip


The work is partially supported by the Ministry of Education, Culture,
Sports, Science and Technology (MEXT), Grant-in-Aid No.19047001,
20041008, 20244015, 20540237, 21340043, 21740140, 22740120, 23000004,
23540265, and Global-COE from MEXT ``The Next Generation of Physics,
Spun from Universality and Emergence'' and ``Nanoscience and Quantum
Physics''.



\onecolumn

\begin{table}\footnotesize
  \begin{center}
    \caption{Source list of AGNs used in our study with spectral and
   luminosity information.}
    \begin{tabular}{rlllrrccrr}
      \hline\hline
No. & MAXI Name & Counterpart & Type & Flux$^a$ & Redshift & $\Gamma$&
     $N_{\rm H}^b$& Log $L_{\rm X}^c$ &Ref.$^d$ \\
\hline
  3& MAXI J0048+320& Mrk 348	& AGN2&  2.42& 0.0150& 1.69& 16& 43.49 & (1)\\
  9& MAXI J0229+315& NGC 931	& AGN1&  1.77& 0.0167& 1.75&  0.36& 43.27 & (1)\\
 18& MAXI J0333--364& NGC 1365	& AGN2&  1.97& 0.0055& 2.11& 29& 42.74 & (2)\\
 24& MAXI J0423--569& 1RXS J042601.6--571202	& AGN1&  1.71& 0.1040& 1.48&  0& 44.85 & (1)\\
 26& MAXI J0433+053& 3C 120	& AGN1&  2.81& 0.0330& 2.00&  0.05& 44.10 & (3)\\
 29& MAXI J0510+166& IRAS 05078+1626	& AGN1&  2.77& 0.0179& 2.07$^{+0.55}_{-0.17}$&  0.0$^{+1.9}$& 43.56 & (4)\\
 31& MAXI J0516--001& Ark 120	& AGN1&  2.24& 0.0323& 1.90&  0& 43.97 & (1)\\
 40& MAXI J0552--075& NGC 2110	& AGN2&  8.13& 0.0078& 1.54&  2.84& 43.28 & (1)\\
 41& MAXI J0555+464& MCG +08-11-011	& AGN1&  3.66& 0.0205& 1.64&  0.25& 43.75 & (1)\\
 49& MAXI J0924--317& 2MASX J09235371--3141305	& AGN2&  5.78& 0.0423& 2.07$^{+0.31}_{-0.17}$&  0.0$^{+0.7}$& 44.64 & (4)\\
 50& MAXI J0947--309& MCG --05-23-016	& AGN2&  6.58& 0.0085& 1.90&  1.6& 43.28 & (1)\\
 52& MAXI J1023+197& NGC 3227	& AGN1&  1.85& 0.0039& 1.52&  6.6& 42.06 & (3)\\
 56& MAXI J1105+725& NGC 3516	& AGN1&  1.84& 0.0088& 1.73&  0.35& 42.73 & (1)\\
 58& MAXI J1139--378& NGC 3783	& AGN1&  3.21& 0.0097& 1.60&  0.09& 43.04 & (3)\\
 61& MAXI J1144--184& 2MASX J11454045--1827149	& AGN1&  1.87& 0.0330& 1.92&  0& 43.91 & (1)\\
 62& MAXI J1210+394& NGC 4151	& AGN1&  6.25& 0.0033& 1.65&  3.4& 42.44 & (3)\\
 67& MAXI J1240--052& NGC 4593	& AGN1&  1.96& 0.0090& 1.69&  0& 42.76 & (3)\\
 75& MAXI J1335--342& MCG --06-30-015	& AGN1&  3.42& 0.0077& 1.92&  0.02& 42.90 & (3)\\
 77& MAXI J1338+045& NGC 5252	& AGN2&  2.93& 0.0230& 1.55&  4.34& 43.80 & (1)\\
 80& MAXI J1349--302& IC 4329A	& AGN1&  7.44& 0.0160& 1.74&  0.36& 43.86 & (3)\\
 83& MAXI J1413--031& NGC 5506	& AGN2&  5.73& 0.0062& 1.72&  3.23& 42.95 & (3)\\
 84& MAXI J1418+251& NGC 5548	& AGN1&  3.06& 0.0172& 1.57&  0& 43.51 & (3)\\
108& MAXI J1716--629& NGC 6300	& AGN2&  1.65& 0.0037& 1.83& 21.5& 42.19 & (1)\\
109& MAXI J1741+185& 4C +18.51	& AGN1&  1.66& 0.1860& 1.90$^{+0.17}_{-0.00}$& 4.0$^{+8.3}_{-4.0}$& 45.47 & (4)\\
115& MAXI J1835+328& 3C 382	& AGN1&  2.69& 0.0579& 1.86&  0.06& 44.56 & (5)\\
117& MAXI J1837--653& ESO 103-035	& AGN2&  2.26& 0.0133& 1.96& 20.3& 43.45 & (3)\\
118& MAXI J1839+798& 3C 390.3	& AGN1&  2.85& 0.0561& 1.64&  0.03& 44.53 & (3)\\
119& MAXI J1851--783& 2MASX J18470283--7831494	& AGN1&  1.45& 0.0741& 1.93&  0.01& 44.53 & (3)\\
122& MAXI J1920--586& ESO 141-G055	& AGN1&  2.52& 0.0360& 1.72&  0& 44.09 & (3)\\
127& MAXI J2009--611& NGC 6860	& AGN1&  1.61& 0.0149& 1.64&  0.10& 43.11 & (6)\\
129& MAXI J2041+750& 4C +74.26	& AGN1&  1.83& 0.1040& 1.86&  0.18& 44.93 & (1)\\
130& MAXI J2044--107& Mrk 509	& AGN1&  3.32& 0.0344& 1.49&  0& 44.15 & (3)\\
132& MAXI J2135--626& 1RXS J213623.1--622400	& AGN1&  1.48& 0.0588& 2.10$^{+0.87}_{-0.20}$&  0.0$^{+3.9}$& 44.35 & (4)\\
135& MAXI J2202--319& NGC 7172	& AGN2&  2.02& 0.0087& 1.69&  8.2& 42.85 & (3)\\
138& MAXI J2235--259& NGC 7314	& AGN2&  2.46& 0.0048& 1.85&  0.72& 42.33 & (3)\\
140& MAXI J2253--177& MR 2251--178	& AGN1&  2.98& 0.0640& 1.41&  0.28& 44.64 & (1)\\
142& MAXI J2305--085& Mrk 926	& AGN1&  3.92& 0.0469& 1.61&  0& 44.51 & (3)\\
      \hline
     \multicolumn{10}{l}{$^a$: observed flux in the 4--10 keV band in
     units of erg cm$^{-2}$ s$^{-1}$}\\
     \multicolumn{10}{l}{$^b$: intrinsic luminosity (before absporption)
     in the 2--10 keV band in
     units of erg s$^{-1}$}\\
     \multicolumn{10}{l}{$^c$: in units of $10^{22}$ cm$^{-2}$}\\
     \multicolumn{10}{l}{\hbox to 0pt{\parbox{170mm}{
     \par\noindent 
$^d$: reference for the X-ray spectra: (1) \citet{win09}, (2) \citet{ris05}, (3) \citet{shi06}, (4)
     MAXI/GSC hardness ratio (this work), (5) \citet{sam11}, (6) \citet{win10}
          }\hss}
     }
      \label{tab:ref_catalog}
    \end{tabular}
  \end{center}
\end{table}

\begin{table}
  \begin{center}
    \caption{Best fit parameters of the \nh\ function and 2--10 keV luminosity
   function of AGNs in the local universe.}
    \begin{tabular}{ccccccc}
      \hline\hline
\multicolumn{3}{c}{\nh\ function}&
\multicolumn{4}{c}{2--10 keV Luminosity Function}\\
\hline
$\epsilon$ & $\Psi_{44}$ & $\beta$ & 
$A^a$ & $\gamma_1$ &$\gamma_2$& $L_{*}^b$ \\
1.3 (fixed) & 0.37$\pm$0.03 & 0.23$\pm$0.02 &
$(1.49\pm0.25)\times 10^{-5}$ 
 &1.00(fixed) &2.13$^{+0.28}_{-0.21}$ & 43.59$^{+0.33}_{-0.41}$ \\
&&&
$(4.2\pm0.7)\times 10^{-5}$ &0.84(fixed) &2.03$^{+0.21}_{-0.17}$ & 43.30$^{+0.31}_{-0.38}$\\
&&&
$(8.8\pm1.5)\times 10^{-5}$ 
 &0.62(fixed) &1.95$^{+0.17}_{-0.14}$ & 43.03$^{+0.28}_{-0.33}$\\
      \hline
      \multicolumn{7}{l}{$^a$: in units of [$h_{\rm 70}^3$ Mpc$^{-3}$]}\\
      \multicolumn{7}{l}{$^b$: in units of [$h_{\rm 70}^{-2}$ \erg]}\\
      \multicolumn{7}{l}{The error is $1\sigma$ for a single parameter.}\\
      \label{tab2}
    \end{tabular}
  \end{center}
\end{table}

\end{document}